\documentclass[a4paper]{article}
\usepackage[margin=25mm]{geometry}
\usepackage{amsmath}
\usepackage{amsfonts}
\usepackage{amssymb}
\usepackage{graphicx}
\usepackage[dvipsnames]{xcolor}
\usepackage[
backend=biber,
style=numeric,
citestyle=ieee,
sorting=none
]{biblatex}

\title{Personal autonomy and surveillance capitalism: possible future developments}
\author{Davide Foini \\ Polytechnic University of Milan \\ davide.foini@mail.polimi.it}
\date{} 

\begin{document}
\maketitle

\begin{abstract}
 The rise of social media and the increase in the computational capabilities of computers have allowed tech companies such as Facebook and Google to gather incredibly large amounts of data and to be able to extract meaningful information to use for commercial purposes. Moreover, the algorithms behind these platforms have shown the ability to influence feelings, behaviors, and opinions, representing a serious threat to the independence of their users. All of these practices have been referred to as "surveillance capitalism", a term created by Shoshana Zuboff. In this paper I focus on the threat imposed on the autonomy of human beings in the context of surveillance capitalism, providing both an analysis of the reasons why this threat exists and what consequences we could face if we take no action against such practices.
\end{abstract} \hspace{10pt}

\section*{Introduction}
\subsection*{Problem presentation}
The last fifteen years have been characterized by the large diffusion of the internet and social media, such as Facebook and Instagram, along with the tendency of users to share their data, both consciously (through posts, photos, etc.), and unconsciously (accepting the terms of service, allowing cookies when navigating the web, etc.).  All this information has become incredibly valuable when coupled with big data practices because that allows companies that hold it to exploit it, extracting new information or behavioral models from it , in order to influence and predict the users' behavior and so to capitalize on the advertisements.\par
Episodes in which this influence has been used outside the logic of business are well-known, from trying to influence the result of an election, like the Cambridge Analytica case \cite{CambridgeAnalytica}, to mass-surveillance, like revealed by Edward Snowden \cite{Snowden}.\par
Therefore it is important to note that threats to our autonomy do not just undermine our integrity as individuals, but are also a serious risk to society as a whole, making it essential to discuss future developments of surveillance capitalism.
\subsection*{Purpose of the paper and definitions}
The purpose of this paper is to focus on the autonomy of individuals and analyze how it could be affected by surveillance capitalism, but before starting the discussion I find it appropriate to introduce some concepts and their definitions.\par
As already mentioned, my discussion revolves around \emph{surveillance capitalism} that was first introduced by Shoshana Zuboff, indicating the increasing ability of capitalism to modify and predict human behavior to increase revenues and control over the market, especially thanks to the new capabilities of information technology \cite{Zuboff2015}. \par
The main concept that my debate is going to revolve around is the one of \emph{autonomy}. Among all the definitions available in the literature I decided to take as reference the definition given by Joseph Raz: "(t)he ruling idea behind the ideal of personal autonomy is that people should make their own lives."\cite[p.~396]{Raz1986-RAZTMO-2}. This involves both being able to take choices regarding one's own life and also to be able to reason about them, taking into account all the personal beliefs and each own background, without being influenced by external factors. \par
Another necessary definition is the one of \emph{manipulation}. There are different ways to define manipulation, but in my opinion, the one that suits the most the purpose of my work is the one given by Susser et al.: "In our view manipulation is hidden influence. Or more fully, manipulating someone means intentionally and covertly influencing their decision-making, by targeting and exploiting their decision-making vulnerabilities."\cite[p.~4]{Susser2019}. The reason why I find this definition so appropriate is that it emphasizes the hidden property of the influence, which is also how surveillance capitalism's mechanisms work, as I am going to point out in one of the next sections. \par
The last concept that I find essential to define is the one of \emph{big data}. When mentioning big data and the correlated mechanisms, I will be referring to the operations of data extraction from all the possible sources, and to the operations performed on the data in order to analyze and extract patterns useful for behavior prediction and therefore manipulation. Big data is essential because it can be considered the turning point for concerns about autonomy: forms of manipulation have long been enforced through traditional media (like newspapers, radio, and lastly via television), but the amount of data gathered with big data enables a "tailored influence", taking it to another order of magnitude of effectiveness. 

\subsection*{Thesis}
I strongly believe that if no action is taken against surveillance capitalism our autonomy will be downsized in the near future. 

\subsection*{Paper structure}
Having expressed my belief, to support it the paper will be structured as follows. In the next sections, I provide some arguments that reinforce my opinion both on a theoretical level and by providing some examples. In the second part, some counterarguments will be presented and I will discredit them through simple reasoning. In the end, I will sum up everything that was said and provide a conclusion for the discussion.\par

\section*{Autonomy as an obstacle to revenue}
The first reason why I believe my concerns about our autonomy in the future are valid is that the big tech companies that benefit from surveillance capitalism will only increase their revenues if they are more able to model, predict and influence our behavior and choices, therefore they will try to reduce our autonomy as much and as fast as possible. This trend comes from the modus operandi of companies in the classic capitalistic market, which can be defined as revenue-driven, as also expressed by Zuboff: "Just as industrial capitalism was driven to the continuous intensification of the means of production, so surveillance capitalists are now locked in a cycle of continuous intensification of the means of behavioral modification" \cite[p.~9]{Zuboff2019}. The following is a straightforward example: Volkswagen would jump at the opportunity to reduce the consumption of their cars in order to improve considerably the performances (and therefore the number of sales and the resulting revenue) as much as Facebook would quickly use a new algorithm that is twice as effective in making us prone to buy a new pair of shoes.\par
Having said this, we can realize that the process of reducing our independence has been going on long before the advent of information technology, but the pace has increased exponentially.

\section*{Social embedding}
The second reason why I think we will face a reduction in our autonomy is that all of the mechanisms that surveillance capitalism uses, meaning the ability to gather incredibly large amounts of data from the users, are embedded in the social tissue through social media, such as Facebook, and smart assistants, such as Amazon Alexa. While smart assistants are still spreading, social media today are used by almost everyone on a daily basis (58.4\% of the world’s total population in January 2022 according to \cite{datareportal}), and are used not only to keep in touch with friends and family but also to read the news and interact with politicians. I think that it is enough to ask ourselves some questions such as: "Could someone live without using social media nowadays?", "Could a politician run a campaign without using Facebook?" and "Could I go somewhere new without using Google Maps?" to realize that the answer to the question "Can we actually choose to be free from being subjects of surveillance capitalism?" is negative.\par
To better understand how much the use of social media is already an important aspect of our social life and can be expected to become more important is the so-called "fear of missing out", better known as FOMO. FOMO can be defined as "a pervasive apprehension that others might be having rewarding experiences from which one is absent" \cite[p.~1]{Przybylski2013}, which has shown the tendency to cause the desire to stay continually connected online. Even though studies (\cite{Przybylski2013, Buglass2017}) have shown how FOMO can be linked with negative effects on one's mood and life satisfaction, this phenomenon highlights the fact that surveillance capitalism mechanisms (in this particular case social media) have become so powerful to influence us psychologically to promote their usage.\par
At this point, it is fair to ask ourselves if the choice of being subject to surveillance capitalism practices is forced upon us or not, and if so, if it can be considered a decision taken with complete autonomy.

\section*{Unawareness}
Another cause for the loss of autonomy that I think we will face due to surveillance capitalism is that most part of individuals is totally unaware, not just of how mechanisms such as cookies and Google advertisements work, but also of their mere existence. It is quite intuitive how this situation is very favorable for companies such as Google and Facebook since they can continue to operate without having to worry about users getting concerned about practices that undermine their autonomy but also because this unawareness has enabled them to extend the level of depth to which our behaviors can be predicted and steered.\par
I think that to better understand how this characteristic of surveillance capitalism could be an important factor in future developments it is useful to analyze a similar problem that we are facing nowadays: climate change. We have seen that even though climate change has been known as a serious danger since the last century, serious action to mitigate the effects has been taken only since the majority of people have learned what it is and has developed a conscience about it. In my opinion, it is the same situation we are facing with surveillance capitalism: as long as common knowledge is not developed there will be no actions against it.\par
A further reason that prevents the creation of collective consciousness is that the mechanisms of surveillance capitalism are shaped in such a way that they are hardly detectable, even for an individual that knows about their existence, as I am going to explain in the next chapter.

\section*{Underlying Functioning}
An important feature is that mechanisms that are used to influence our behavior and choices, for example, what is shown in our Facebook feed or which articles Amazon recommends to us, are fed with data that the user generates without being conscious of it (an example could be the area on the screen of the smartphone that is touched) and that these algorithms work without the users noticing.\par
Information technology has already gone "in the background", meaning that besides being socially embedded, as I explained previously, we use it without noticing it: it has become a natural way of behaving for certain operations. For example, it is quite natural, when hearing about a person we don't know, to automatically look for her/his profile on social media or check the reviews of a restaurant on Google or Trip Advisor before going out for dinner.\par
To better understand the level of this underlying functioning Susser et al. defined digital platforms as eyeglasses instead of magnifying glasses, meaning that our attention is focused on the information that technologies provide us (like videos, photos, or directions to follow) instead of the technology itself \cite{Susser2019-2}. Following this realization, we can see how the aim of the big tech companies that profit from surveillance capitalism practices is to develop technologies that tend to become invisible even when they are actually in our hands, like smartphones and smartwatches, or inside our houses, like smart assistants. All of this is another obstacle to the creation of collective consciousness that will be necessary to regulate surveillance capitalism. \par

\section*{Limitless Reaching}
A further aspect that can be underestimated is that technologies that are used to gather useful data and through which we "look" at the world, as seen in the previous section, are everyone around us and in all aspects of everyday life. This gives surveillance capitalism an unlimited scope of action and furthermore makes it more difficult for us to avoid being subjects of both data gathering and manipulation, giving us little or no actual space when we are completely free from being influenced. In this sense, our smartphones represent the first way of gathering data about us: we take them with us all of the time and they are able to know our location, they have access to our audio to detect if we are calling the smart assistant, and they have become the filter through which we interact with the world.\par 
The scope of surveillance capitalism does not end with everything that can be gathered from us, but it also extends to what can be derived from the data thanks to big data analyses. An example of future developments is the insurance sector: gathering data from our cars, like the way we drive and where we drive, big data would be able to understand how much will be the chance of us getting into a car accident, and therefore the insurance will be able to require a higher fee even if we always respect the traffic regulations. Other information, like the ethical group, could be derived from data such as the type of videos watched and could be taken into account for an insurance policy.\par
Another growing concern is that this scope, which already includes our social and private life, is also starting to involve the workplace.
An example is the wristbands that Amazon patented for warehouse workers that via haptic signals can point the wearer to the right products and keep track of the position of the workers' hands. Allegedly the main reason for such devices is to save labor time, but it is intuitive to see how this mechanism could be used to steer the hand movements of the employees to avoid "useless gestures" like scratching and impose a pace of operations. Fortunately, the news (\cite{Amazon1, Amazon2}) of this patent attracted a lot of attention and the bracelets are still only patented and not put into operation (as of today).\par
So far I have defended my claim with different arguments, and in the next sections, I will present possible counterarguments, illustrating why they do not discredit my opinion.

\section*{Free influence or expensive autonomy?}
A possible reason for supporting surveillance capitalism practices is that they are the reason why, nowadays, we have access to an enormous variety of services and content for free (not taking into account the cost of having an internet connection, which is negligible and has been diminishing since its beginning) or products for a very moderate cost (like smart assistants). Therefore it would be reasonable to give up our data for such a bargain.\par Social media enable us to instantly connect with potentially everyone in the world, sharing messages, pictures, and videos without a fee, unlike phone calls and SMSs or MMSs. Information is also accessible everywhere, anytime, and of every type just with a "click" or a "tap", when before it was necessary to go to a library or to buy a newspaper. It is undeniable that being free represents a point in favor at first sight, but are they actually free?\par
First, I think that it is plain to see that data has become a high-value strategic asset: the wealth of big tech companies originates from it, therefore it is wrong to argue that those services are free, we just give back something of value that is not money.\par
Furthermore, as seen so far, data is just the starting point because it is not only used for general purposes but can also be put up against us to undermine our autonomy. An example is online shopping: if we are interested in a new smartphone, not just because we actually need it but maybe just out of curiosity, we search on Amazon and we instantly get a list of different items from various brands, we can then compare prices, the specifications, and also read the reviews from other users. While we get all of this, Amazon keeps track of our research and sells this information to third parties, and in the next period, our Facebook home page will start to contain posts about smartphone sales and we will end up buying a new one.\par
To sum up, we are not giving up mere information about ourselves (and all the data that can be inferred from it), but also pieces of ourselves, that can be our opinions, beliefs, and tastes just to name a few. These are the reasons why everything that is promoted as free in this field has a twofold price truthfully: our data and our autonomy.

\section*{Giving up autonomy for our own sake}
Another argument against the limitations of surveillance capitalism practices is that the amount of data that is gathered is so huge and can unlock a knowledge so deep about us that it is possible to influence us to act for what is perceived as our own good. An example in this sense can be a mobile application that can promote a healthy lifestyle, encouraging exercising and a healthy diet. In this case, manipulation would be perceived with the welfare of the subject as objective as a goal.\par
To understand why this reasoning is flawed it is enough to point out that "our own good" must be defined by someone, resulting in a paternalistic model, where that someone "takes the wheel" for us because he is better at it. What is most concerning is that this paradigm could be extended not just to matters related to our physical well-being, but also to all the aspects of our daily lives, like a smart device that tracks the energy usage of a household adjusting the heating and cooling systems accordingly. This would mean giving up the freedom to make the wrong decisions, such as having a habit like smoking cigarettes or  deciding to buy the car we most like and not because it would be the best for us in terms of consumption, range, etc.\par
It is also essential to remember that all the data and the power of prediction and manipulation are in the hands of private companies, whose goal is mere profit and not our interests. In the two examples cited before the company owning the mobile application or the smart device would sell the data collected (our habits or data about the dimension of our house) to third parties, emphasizing the lack of reasons to trust such organizations with our autonomy. 

\section*{Conclusion}
In this paper, I argued that if we don't take measures against surveillance capitalism, we will be subjected to a reduction of our autonomy. In the first part, I have explained why I fear this is going to happen. The first reason is that such mechanisms follow the capitalistic models and therefore aim at becoming more efficient in influencing and manipulating us. Moreover, they have a pervasive presence and we are not able to escape them. They also work without being noticed and without hardly anyone knowing how they actually work, making it very hard to create a collective consciousness that would be necessary to take action against such mechanisms.\par
In the second part, I discussed possible reasons to be in favor of surveillance capitalism or against restricting it. The first reason treated is the fact that it is thanks to surveillance capitalism that we have access to so many contents and services for free, but I have shown that is intrinsically wrong to define those services as "free" since we don't pay with money but with our data, that is sold to third parties and it is also used with the scope of influencing us and undermining our autonomy, therefore having a doubled price. The second idea was the one that getting influenced could be for our own good, but I have pointed out that this is dangerous since what is considered good for us is to be defined by private companies, whose objective is ultimately profit and for sure not our welfare, therefore raising concerns about willingly giving them the keys to our freedom.\par
In conclusion, it is fair to say that the current situation with regard to surveillance capitalism practices, like massive data gathering, behavioral prediction, and manipulation, already rises a lot of concerns for our autonomy, and it is expected to get worse, following the trend that it has followed since the early days if no action is taken against such techniques. Therefore to tackle these problems it is necessary to act both on a social and a policy level, otherwise, the consequences are already before our very own eyes.
\printbibliography

@article{Zuboff2015,
author = {Zuboff, Shoshana},
doi = {10.1057/jit.2015.5},
issn = {14664437},
journal = {Journal of Information Technology},
number = {1},
pages = {75--89},
title = {{Big other: Surveillance capitalism and the prospects of an information civilization}},
volume = {30},
year = {2015}
}

@article{Zuboff2019,
author = {Zuboff, Shoshana},
doi = {10.1177/1095796018819461},
isbn = {1095796018819},
issn = {15572978},
journal = {New Labor Forum},
number = {1},
pages = {10--29},
title = {{Surveillance Capitalism and the Challenge of Collective Action}},
volume = {28},
year = {2019}
}

@book{Raz1986-RAZTMO-2,
	author = {Joseph Raz},
	year = {1986},
	title = {The Morality of Freedom},
	publisher = {Oxford University Press}
}

@article{Susser2019,
author = {Susser, Daniel and Roessler, Beate and Nissenbaum, Helen},
doi = {10.14763/2019.2.1410},
issn = {21976775},
journal = {Internet Policy Review},
number = {2},
pages = {1--22},
title = {{Technology, autonomy, and manipulation}},
volume = {8},
year = {2019}
}

@online{datareportal,
    author = "Simon Kemp",
    title = "Digital 2022: Global Overview Report",
    url  = "https://datareportal.com/reports/digital-2022-global-overview-report",
    urldate  = {2022-03-12}
}

@article{Susser2019-2,
author = {Susser, Daniel and Roessler, Beate and Nissenbaum, Helen F.},
doi = {10.2139/ssrn.3306006},
journal = {SSRN Electronic Journal},
pages = {1--45},
title = {{Online Manipulation: Hidden Influences in a Digital World}},
volume = {1},
year = {2019}
}

@online{CambridgeAnalytica,
    author = "Al Jazeera News",
    title = "Cambridge Analytica and Facebook: The scandal so far",
    url  = "https://www.aljazeera.com/news/2018/3/28/cambridge-analytica-and-facebook-the-scandal-so-far",
    urldate = {2023-07-01}
}

@online{Amazon1,
    author = "Olivia Solon",
    title = "Amazon patents wristband that tracks warehouse workers' movements",
    url  = "https://www.theguardian.com/technology/2018/jan/31/amazon-warehouse-wristband-tracking",
    urldate = {2023-04-01}
}

@online{Amazon2,
    author = "Ceylan Yeginsu",
    title = "If Workers Slack Off, the Wristband Will Know. (And Amazon Has a Patent for It.)",
    url  = "https://www.nytimes.com/2018/02/01/technology/amazon-wristband-tracking-privacy.html",
    urldate = {2023-04-01}
}

@online{Snowden,
    author = " Mark Mazzetti, Michael S. Schmidt",
    title = "Ex-Worker at C.I.A. Says He Leaked Data on Surveillance",
    url  = "https://www.nytimes.com/2013/06/10/us/former-cia-worker-says-he-leaked-surveillance-data.html",
    urldate = {2023-07-01}
}

@article{Buglass2017,
author = {Buglass, Sarah L. and Binder, Jens F. and Betts, Lucy R. and Underwood, Jean D.M.},
doi = {10.1016/j.chb.2016.09.055},
issn = {07475632},
journal = {Computers in Human Behavior},
pages = {248--255},
publisher = {Elsevier Ltd},
title = {{Motivators of online vulnerability: The impact of social network site use and FOMO}},
url = {http://dx.doi.org/10.1016/j.chb.2016.09.055},
volume = {66},
year = {2017}
}

@article{Przybylski2013,
author = {Przybylski, Andrew K. and Murayama, Kou and Dehaan, Cody R. and Gladwell, Valerie},
doi = {10.1016/j.chb.2013.02.014},
issn = {07475632},
journal = {Computers in Human Behavior},
number = {4},
pages = {1841--1848},
publisher = {Elsevier Ltd},
title = {{Motivational, emotional, and behavioral correlates of fear of missing out}},
url = {http://dx.doi.org/10.1016/j.chb.2013.02.014},
volume = {29},
year = {2013}
}
\end{document}